# Characterisation of Li in the surface film of a corrosion resistant Mg-Li(-Al-Y-Zr) alloy

Y. Yan[a], Y. Qiu[a], O. Gharbi[b], N. Birbilis*[,c] and P.N.H. Nakashima[a]

[a]*Department of Materials Science and Engineering, Monash University, VIC 3800, Australia*
[b]*Sorbonne Université, CNRS, Laboratoire Interfaces et Systèmes Electrochimiques, LISE, F-75005 Paris, France*
[c]*College of Engineering and Computer Science, Australian National University, Acton, ACT 2601, Australia*

*Corresponding author: nick.birbilis@anu.edu.au (N. Birbilis)

## Abstract

The surface film formed upon Mg-Li(-Al-Y-Zr) following aqueous immersion and air-exposure was investigated. This alloy (which contains 30.3 at. % Li) possesses a bcc crystal structure and has been reported as being corrosion resistant. It was determined that the principal components of the surface film were $Li_2CO_3$ and $Mg(OH)_2$ as characterised by grazing incidence X-ray diffraction (GIXRD). The detection of hcp grains near the alloy surface was observed by GIXRD and selected area electron diffraction (SAED). The spatial distribution of Li and Mg in the surface film was characterised by electron energy loss spectroscopy (EELS) and the distribution of other major elements in the alloy was characterised by scanning transmission electron microscopy (STEM) and energy dispersive X-ray spectroscopy (EDXS). It was observed that Li was distributed throughout the alloy surface film and with an elevated concentration in the so-called outer layer.

# 1. Introduction

Magnesium-lithium (Mg-Li) alloys are a promising ultra-lightweight class of alloys for applications where material density is a critical factor [1-3]. By appropriate alloy design and processing, Mg-Li alloys can achieve high specific strength, whilst the addition of Li beyond ~5.7 wt. % can markedly improve alloy ductility by altering the crystal structure of the alloy matrix from hexagonal close-packed (hcp) to body-centred cubic (bcc) [4-7]. When the addition of Li exceeds 10.3 wt. %, single phase bcc Mg-Li forms, resulting in high ductility coupled with low alloy density [2, 7].

The influence of Li additions on the corrosion of Mg alloys has been studied by numerous groups [5, 8-11]. It has been demonstrated that bcc Mg-Li alloys (where Li > 10.3 wt. %) have superior corrosion resistance to Mg-Li alloys with lower Li concentration. This has been attributed to the combined effect of the formation of a uniform bcc matrix which eliminates galvanic coupling between the hcp α-Mg and bcc β-Li phases, and more critically, the formation of a protective $Li_2CO_3$ surface film upon bcc Mg-Li [2, 8].

The formation of $Li_2CO_3$ (or $Li_2O$) containing surface films upon Mg-Li alloys has been reported by several independent groups to date [2, 10-12]. However, the spatial or physical characterisation of Li in the surface film of Mg-Li alloys is still challenging. In a previous study, by mapping the distribution of O and C (in addition to mapping Mg and visualising its depletion) using energy dispersive X-ray spectroscopy (EDXS), a thin layer of $Li_2CO_3$ was inferred on top of the surface film of an Mg-Li(-Al-Y-Zr) alloy [13]. However, the distribution of Li or such Li compounds in the surface film of Mg-Li alloys is still yet to be revealed.

To spatially characterise Li in the surface film of an Mg-Li(-Al-Y-Zr) alloy, electron energy loss spectroscopy (EELS) was used in the present work. EELS is a technique that has been widely applied to physically and spatially characterise the Li transition metal oxides in the study of lithium-ion batteries [14-16]. The EELS technique can quantitatively (under certain conditions) characterise elemental distributions even for the light elements such as Li, which are not possible to detect using EDXS [17, 18]. Although the EELS technique is not widely used for surface characterisation of light alloys, EELS mapping has been previously attempted to characterise Li in aluminium alloys [19]. Herein, EELS mapping on a cross-sectionally prepared lamella of a bcc Mg-Li(-Al-Y-Zr) alloy was carried out in order to determine the distribution of Li and Mg in the surface film.

## 2. Material and methods

*2.1 Materials*

The Mg-Li(-Al-Y-Zr) alloy investigated in the present work has the following composition, Mg - 30.3 Li - 2.34 Al - 0.128 Y - 0.039 Zr (in at. %). The alloy was prepared by gravity casting, then homogenised, extruded, heat treated and water quenched, artificially aged and cold-rolled, following the previously described process [2]. Alloy sheet was cut into specimens of ~1 × 1 cm$^2$ and successively ground with SiC paper under ethanol to a 2000 grit finish, then rinsed with ethanol. The test specimens were then immersed in quiescent 0.1 M NaCl for 24 hours, rinsed with ethanol and exposed to atmospheric conditions for 7 days prior to characterisation.

*2.2 Grazing incidence X-ray diffraction (GIXRD)*

The surface film of the immersed then air-exposed sample was investigated using a Bruker D8 Discover X-ray diffractometer operating with a copper target (K$_α$ = 1.5406 Å) at 40 kV and 40 mA coupled with a high energy resolution Lynxeye XE-T detector. The grazing incidence X-ray diffraction (GIXRD) mode with a grazing angle of 0.2° was used to obtain the composition of the surface film. Angles of incidence of 1° and 5° were also used to characterise the crystal structure of the matrix near the surface film. All of the spectra were collected within a 2θ range of 15° to 63°.

*2.3 Analytical electron microscopy*

The elemental distribution within the surface film was characterised using EELS and EDXS. The TEM lamella for the tests was prepared using a focused ion beam (FIB) in an FEI Quanta 3D FEG SEM. A cross-sectional lamella with a thickness of 1 µm was milled from the immersed then air-exposed surface of the alloy using a Ga ion beam. Through micro-manipulation, the lamella was removed and "welded" to a Cu grid. The lamella was subsequently milled and fine-polished using a low energy Ga ion beam to a final thickness of ~100 nm for TEM characterisation.

Bright field (BF) and high angle annular dark field (HAADF) images were collected using an FEI F20 TEM operating at 200 kV. This instrument also permitted the collection of EDXS maps of Mg, Al, C and O within the surface film.

The EELS characterisation was performed using a JEOL 2100F TEM, operating at 200 kV, and a post-column Gatan Enfina EEL spectrometer. Spectrum imaging [20 – 24] was carried out with a nominal probe size of 0.5 nm and an inter-pixel distance of 5 nm. Two different areas on the TEM lamella specimen were interrogated, the first spanning an area of 397 nm x 630 nm and the second, an area of 354 nm x 802 nm. The second of these dimensions is perpendicular to the alloy surface (extending in the depth direction). Both areas subjected to spectrum imaging spanned the "platinum deposition layer", "outer layer", "inner layer" and "matrix" of the alloy surface region. Note that the EELS spectrum images and EDXS maps were obtained from different parts of the surface film owing to the beam damage caused throughout both types of data acquisition.

Analysis of the EEL spectrum images required a sensitive method for separating the signals due to Mg (which has a broad $L_{2,3}$ peak with an onset at $\Delta E = 51$ eV and multiple maxima) and the signal due to Li (which has a sharp K edge with an onset at $\Delta E = 55$ eV and a single maximum) [25 – 27], which overlap in a non-trivial fashion when these two elements are present in a material. In summary, this separation was achieved by the following steps for each spectrum image:

(i) Background subtraction was carried out using the pre-Mg $L_{2,3}$ energy loss region of $\Delta E = 44 \pm 4$ eV to determine the background function.

(ii) All of the background-subtracted spectra in the spectrum image were integrated to obtain an "average spectrum" representative of the whole region being interrogated.

(iii) Each individual background-subtracted spectrum in a spectrum image was divided by the "average spectrum" determined in (ii) to produce a spectrum "ratio image" in which spectral prominences indicate elevated concentrations of the corresponding element at the position in question, relative to the average concentration across the entire region of interest.

(iv) Elemental maps were obtained by centring ±2 eV windows on energy losses associated with prominences unique to Mg and Li in the spectrum "ratio image" and integrating over these windows. This was followed by subtracting a "background ratio" determined on either side of each bracketed prominence, using the same energy window width but with centrings symmetrically displaced with respect to the prominence-bracketing window.

(v) The resultant elemental "redistribution maps" (owing to the fact that the ratios are relative to the average composition of the region being interrogated) for Mg and Li were normalised such that their sum would as closely reproduce the fully integrated, as-captured, spectrum image, given the latter is essentially a thickness map. This approximation is only valid to first order if the elements being mapped make up the majority of the region of the specimen being examined. In the present case, Mg and Li together make up more than 90% of the total alloy composition, placing this alloy within the range of validity of the elemental mapping technique described above.

# 3. Results & Discussion

*3.1 Composition of the surface film and crystal structure of the matrix near the surface*

Typical GIXRD profiles collected at different angles of incidence from the surface of Mg-Li(-Al-Y-Zr) after 24 hours of immersion in 0.1 M NaCl and 7 days of air-exposure are shown in Fig. *1*.

When the angle of incidence was small (0.2°), the X-rays did not penetrate the surface film sufficiently and only signals from $Mg(OH)_2$, as well as a weaker signal from $Li_2CO_3$, were identified in the XRD profile. Other products may also be present in the surface film, but too low in volume fraction to be detected (e.g. Al, Y and Zr-containing products or Mg hydro-carbonates). In our previous research on the same Mg-Li(-Al-Y-Zr) alloy immersed for only 4 hours [13], the GIXRD profile collected at an angle of incidence of 0.2° showed a very intense matrix signal, which may indicate that the surface film has considerably thickened after 24 hours of immersion.

When the angle of incidence was increased to 1° and 5°, the GIXRD profiles started to show signals collected from the matrix. Weak signals from $Al_2Y$ were observed and attributed to the $Al_2Y$ dispersoids in the alloy which serve as grain refiners. As observed by Xu et al.[2], this Mg-Li(-Al-Y-Zr) alloy has a single phase bcc β-matrix, which should not contain any α phase owing to the high Li concentration in this alloy. However, intense signals from hcp α-Mg phase were also observed when the angles of incidence were 1° and 5°. Moreover, compared with the XRD profile collected at 1°, the signal from the bcc β-Li phase increased, while the signal from the hcp α-Mg phase decreased when the angle of incidence was increased to 5°. Since increasing the angle of incidence effectively increases the penetration depth of the X-rays, the change in the signal intensity suggests the presence of a heterogeneous distribution of α-Mg phase in the alloy where α-Mg grains are preferentially distributed near the surface.

Microstructural observations of the matrix in the vicinity of the surface film were carried out by using BF-TEM and SAD as shown in Fig. 2 . The structure of the imaged surface cross-section can be divided into 4 major sections: the top Pt deposited layer created by the FIB sample "lift-out" procedure, the porous outer layer of the surface film, the inner layer of surface film and the matrix. The effect of the cold-rolling process can be observed from the presence of large amounts of relatively small and elongated grains locate under the surface film. As demonstrated in Fig. 2 (b) and (c), the existence of hcp grains near the surface film is confirmed, which corroborates the observation of GIXRD. Since the α phase is a Li-lean phase compared with β phase, the presence of α-Mg grains near the surface is the manifestation of the depletion of Li near the surface which suggests a selective dissolution of Li happened during immersion. As Li concentration in the Mg-Li(-Al-Y-Zr) alloy is only about 10.95 wt. % (30.3 at. %), the selective dissolution of Li near the surface might decrease the local Li concentration below the threshold for the bcc phase (around 10.3 wt. %), thus enabling the formation of α grains. Therefore, it is proposed that the preferential distribution of hcp α grains near the surface was the result of a local phase transformation driven by the selective dissolution of Li. The formation of hcp α grains on the surface may disrupt the electrochemical homogeneity of the alloy and lead to galvanic coupling between the α and β grains which can severely compromise the corrosion resistance of the alloy. Therefore, understanding the formation process and accurate distribution of the α phases in the Mg-Li alloys would be critical. However, since the average grain size in this cold-rolled alloy is relatively small, it is difficult to verify the crystal structures of large numbers of grains using SAED in order to determine the change of the crystal structures as a function of depth. Further characterisation like transmission electron backscatter diffraction (T-EBSD) may be required in future.

*3.2 Elemental distribution in the surface film*

STEM-EDXS maps were collected to reveal the distribution of Mg, Al, O and C in the surface film (Fig. 3). It was observed that the C and O signals are present throughout the surface film which may suggest the ubiquitous formation of carbonates in the surface regions. Notably, there is an apparent higher concentration of C in the top half of the "inner layer" while O is more concentrated in the lower half of this region. The difference in the distribution of C and O may suggest the preferential formation of carbonates in the top half of the inner layer of the surface film. Since the Mg and Al signals are present throughout the inner layer, without direct observation of Li, it is difficult to conclude whether the strong signal from C is associated with $Li_2CO_3$ or Mg carbonate/hydro-carbonates (or even small amounts of Al hydro-carbonates).

Due to the very low excitation energies of low atomic number elements, EDXS cannot detect such elements, including Li, with any statistical confidence. Therefore, to directly characterise the distribution of Li, EEL spectrum imaging was used to map the distribution of both Li and Mg within the surface material of the alloy, according to the methodology described in the previous section. Two different regions within the same TEM lamella specimen were mapped and the results are presented in Fig. 4.

From the Li maps of both regions shown in Fig. 4, it is evident that the Li is concentrated more strongly in the outer layer of the surface region and is uniformly distributed elsewhere. By extracting the average signal intensity along the marked area (inside the yellow box in Fig. 4), semi-quantitative Mg and Li distribution profiles as a function of depth were obtained. It can be observed that Mg is more concentrated in the inner layer of the surface film while there is a higher concentration of Li in the outer layer. In previous research, the outer layer of the surface film formed upon Li-free Mg and Mg alloys was observed to be relatively porous and filled with platelet $Mg(OH)_2$ particles as a result of the continuous dissolution-precipitation of $Mg(OH)_2$ on the surface [28]. However, notwithstanding the needle-shaped $Mg(OH)_2$, other corrosion products which are more evident in the BF images, may also be observed throughout the outer layer of the surface film around these needle-shaped particles as shown in the BF images in Fig. 2 and Fig. 3. From the EDXS maps in Fig. 3, it is evident that the corrosion products found around the needle-shaped particles are rich in C, O and poor in Mg. Therefore, the high Li content observed throughout the outer layer of the surface film by EELS may suggest the relatively transparent corrosion product found throughout the outer layers of the surface film by BF-TEM, is principally $Li_2CO_3$. This corroborates the previous XPS observation that indicates the presence of a $Li_2CO_3$-rich outer layer in the surface film of Mg-Li alloys [2, 11]. In previous research, the formation of $Li_2CO_3$ surface film was proposed to be able to protect the underlying matrix, helping to improve the corrosion resistance of Mg-Li alloys [2]. However, the protective ability of this $Li_2CO_3$-rich outer layer may be limited owing to the porous nature of this layer as observed in the BF-TEM images.

On the other hand, the inner layer of the surface film, which is enriched in Mg, is seen to be relatively dense compared to the outer layer. From the EDXS elemental maps, it can be seen that the Mg signals from the matrix are much stronger than the Mg signals from the surface film. However, part of the matrix shows relatively weak Mg signals in Fig. *4*, which may be attributed to the damage caused by the Ga ion beam during FIB milling of the sample. From previous research regarding the surface film of Mg-Li alloys, the Li was mainly suggested to be distributed at the outer part of the surface film [2, 11-13]. However, the work presented herein has elucidated that in addition to the top $Li_2CO_3$ film, Li signals were also observed throughout the inner layer, overlapping with the signals from Mg and Al. Therefore, comparing with the previously proposed uniform $Li_2CO_3$ film, it appears that a much more complex surface evolves. This is supported by the observation of Al-containing Li-LDH on the surface of the Mg-Li(-Al-Y-Zr) alloy in a previous study [13]. Other Li-containing products may also be present, which supports the need for dedicated characterisation of Li species. Since the protective ability of the porous $Li_2CO_3$ outer layer may not be the sole cause for the appreciable corrosion resistance of this alloy, further research for a more detailed understanding of the inner layer of the surface film of Mg-Li-(Al-Y-Zr) alloys is warranted.

## 4. Conclusion

The present work sought to physically study and characterise the surface film formed upon a corrosion resistant Mg-Li(-Al-Y-Zr) alloy, after immersion in aqueous solution and exposure to air under atmospheric conditions. The comprehensive knowledge of the surface film is essential for understanding the origins of corrosion resistance in this class of Mg-Li alloys. The results herein were both comprehensive and significantly reinforced by the ability of electron energy loss spectrum imaging to spatially map the distribution of Li in the surface film of the alloy. The crystal structure of the matrix grains in the vicinity of the surface film was determined by GIXRD and SAED. The results revealed the presence of an hcp structure in contrast to the typical bcc structure of the bulk. The origin of the hcp structure was posited to be due to the selective dissolution of Li during exposure to the aqueous electrolyte, altering the near-surface chemistry and thus crystal structure. The elemental distribution in the surface film was characterised using EELS and STEM-EDXS. It was observed that the outer layer of the surface film was rich in $Li_2CO_3$. However, the outer layer was observed to be relatively porous. A qualitatively lower concentration, albeit enriched relative to the underlying alloy, of Li was also evident in the inner layer of the surface film - mingled with other surface products to form a relatively dense layer. The EELS method adopted herein was able to provide the spatial (but not quantitative) distribution of Li, whilst comparative Li and Mg depth profiles were also determined, nonetheless providing a significant advance in the ability to interpret the spatial distribution of Li in metallic alloys. Such a characterisation capability has broad consequences in the study of many alloy systems.

## Acknowledgements

We gratefully acknowledge the Monash Centre for Electron Microscopy (MCEM), and the Monash X-ray Platform (MXP). We also thank Dr. Jisheng Ma for technical assistance with GIXRD. We also thank Prof. Mike Ferry and Dr. Martin Xu at UNSW. This work was funded by the Australian Research Council.

# References

[1] M. Esmaily, J.E. Svensson, S. Fajardo, N. Birbilis, G.S. Frankel, S. Virtanen, R. Arrabal, S. Thomas, L.G. Johansson, Fundamentals and advances in magnesium alloy corrosion, Prog. Mat. Sci. 89 (2017) 92-193.

[2] W. Xu, N. Birbilis, G. Sha, Y. Wang, J.E. Daniels, Y. Xiao, M. Ferry, A high-specific-strength and corrosion-resistant magnesium alloy, Nat. Mater. 14 (2015) 1229-1235.

[3] H. Haferkamp, R. Boehm, U. Holzkamp, C. Jaschik, V. Kaese, M. Niemeyer, Alloy Development, Processing and Applications in Magnesium Lithium Alloys, Mater. Trans. 42 (2001) 1160-1166.

[4] Z. Zeng, N. Stanford, C.H.J. Davies, J.-F. Nie, N. Birbilis, Magnesium extrusion alloys: a review of developments and prospects, Int. Mater. Rev. 64 (2019) 27-62.

[5] K. Gusieva, C.H.J. Davies, J.R. Scully, N. Birbilis, Corrosion of magnesium alloys: the role of alloying, Int. Mater. Rev. 60 (2015) 169-194.

[6] A.A. Nayeb-Hashemi, J.B. Clark, Phase diagrams of binary magnesium alloys, ASM International, Materials Park, OH, 1988.

[7] S. Tang, T. Xin, W. Xu, D. Miskovic, G. Sha, Z. Quadir, S. Ringer, K. Nomoto, N. Birbilis, M. Ferry, Precipitation strengthening in an ultralight magnesium alloy, Nat. Commun. 10 (2019) 1003.

[8] C.Q. Li, D.K. Xu, X.-B. Chen, B.J. Wang, R.Z. Wu, E.H. Han, N. Birbilis, Composition and microstructure dependent corrosion behaviour of Mg-Li alloys, Electrochim. Acta 260 (2018) 55-64.

[9] T. Morishige, Y. Obata, T. Goto, T. Fukagawa, E. Nakamura, T. Takenaka, Effect of Al Composition on the Corrosion Resistance of Mg-14 mass% Li System Alloy, Mater. Trans. 57 (2016) 1853-1856.

[10] L. Hou, M. Raveggi, X.-B. Chen, W. Xu, K.J. Laws, Y. Wei, M. Ferry, N. Birbilis, Investigating the Passivity and Dissolution of a Corrosion Resistant Mg-33at.%Li Alloy in Aqueous Chloride Using Online ICP-MS, J. Electrochem. Soc. 163 (2016) C324-C329.

[11] R.C. Zeng, L. Sun, Y.F. Zheng, H.Z. Cui, E.H. Han, Corrosion and characterisation of dual phase Mg-Li-Ca alloy in Hank's solution: The influence of microstructural features, Corros. Sci. 79 (2014) 69-82.

[12] Y. Song, D. Shan, R. Chen, E.H. Han, Investigation of surface oxide film on magnesium lithium alloy, J. Alloys Compd. 484 (2009) 585-590.

[13] Y. Yan, N. Birbilis, O. Gharbi, A. Maltseva, X. Chen, Z. Zeng, S. Xu, W. Xu, P. Volovitch, M. Ferry, Investigating the structure of the surface film on a corrosion resistant Mg-Li(-Al-Y-Zr) alloy, Corrosion (2018).

[14] R. Huang, Y. Ikuhara, STEM characterization for lithium-ion battery cathode materials, Current Opinion in Solid State and Materials Science 16 (2012) 31-38.

[15] J.-M. Chen, C.-H. Hsu, Y.-R. Lin, M.-H. Hsiao, G.T.-K. Fey, High-power LiFePO4 cathode materials with a continuous nano carbon network for lithium-ion batteries, Journal of Power Sources 184 (2008) 498-502.

[16] J. Kikkawa, T. Akita, M. Tabuchi, M. Shikano, K. Tatsumi, M. Kohyama, Fe-rich and Mn-rich nanodomains in Li1.2Mn0.4Fe0.4O2 positive electrode materials for lithium-ion batteries, Appl. Phys. Lett. 91 (2007) 054103.

[17] D.R. Liu, D.B. Williams, Accurate quantification of lithium in aluminium-lithium alloys with electron energy-loss spectrometry, Proceedings of the Royal Society of London. A. Mathematical and Physical Sciences 425 (1989) 91-111.

[18] R. Brydson, A brief review of quantitative aspects of electron energy loss spectroscopy and imaging, Materials Science and Technology 16 (2000) 1187-1198.

[19] P. Visser, Y. Liu, X. Zhou, T. Hashimoto, G.E. Thompson, S.B. Lyon, L.G.J. van der Ven, A.J.M.C. Mol, H.A. Terryn, The corrosion protection of AA2024-T3 aluminium alloy by leaching of lithium-containing salts from organic coatings, Faraday Discuss. 180 (2015) 511-526.

[20] J.-L. Lavergne, J.-M. Martin, M. Belin, Interactive electron energy-loss elemental mapping by the "Imaging-Spectrum" method, Microsc. Microanal. Microstruct. 3 (1992), 517-528.

[21] J. Mayer, U. Eigenthaler, J.M. Plitzko, F. Dettenwanger, Quantitative analysis of electron spectroscopic imaging series, Micron 28 (1997), 361-370.

[22] P.J. Thomas, P.A. Midgley, Image-spectroscopy – I. The advantages of increased spectral information for compositional EFTEM analysis, Ultramicroscopy 88 (2001), 179-186.

[23] P.J. Thomas, P.A. Midgley, Image-spectroscopy – II. The removal of plural scattering from extended energy-filtered series by Fourier deconvolution, Ultramicroscopy 88 (2001), 187-194.

[24] G. Botton, Analytical electron microscopy, Science of Microscopy Volume 1, P.W. Hawkes, J.C.H. Spence Eds (Springer Science + Business Media, New York, 2007), 273-405.

[25] V. Mauchamp, P. Moreau, G. Ouvrard, F. Boucher, Local field effects at Li K edges in electron energy-loss spectra of Li, $Li_2O$ and LiF, Phys. Rev. B 77 (2008), 045117.

[26] J.K. Okamoto, D.H. Pearson, A. Hightower, C.C. Ahn, B. Fultz, EELS analysis of the electronic structure and microstructure of metals, Transmission electron energy loss spectrometry in materials science and the EELS atlas 2$^{nd}$ Ed., C.C. Ahn Ed. (Wiley-VCH, Weinheim, Germany 2004), 317-352.

[27] A. Feldhoff, E. Pippel, J. Woltersdorf, Interface engineering of carbon-fibre reinforced Mg-Al alloys, Advanced Engineering Materials 2 (2000), 471-480.

[28] J.H. Nordlien, S. Ono, N. Masuko, K. Nisancioglu, A TEM investigation of naturally formed oxide films on pure magnesium, Corros. Sci. 39 (1997) 1397-1414.

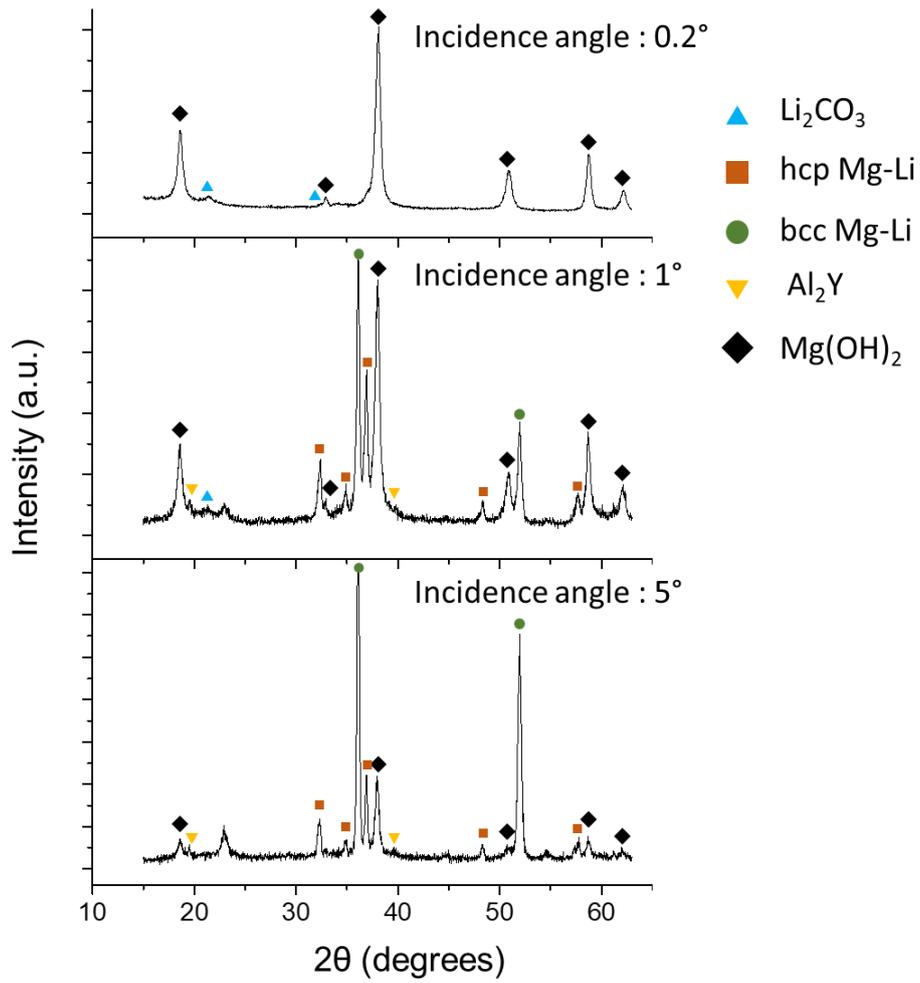

Fig. 1. GIXRD profiles collected with different angles of incidence from the Mg-Li(-Al-Y-Zr) alloy following 24 h of immersion in 0.1 M NaCl and exposure to air for 7 days.

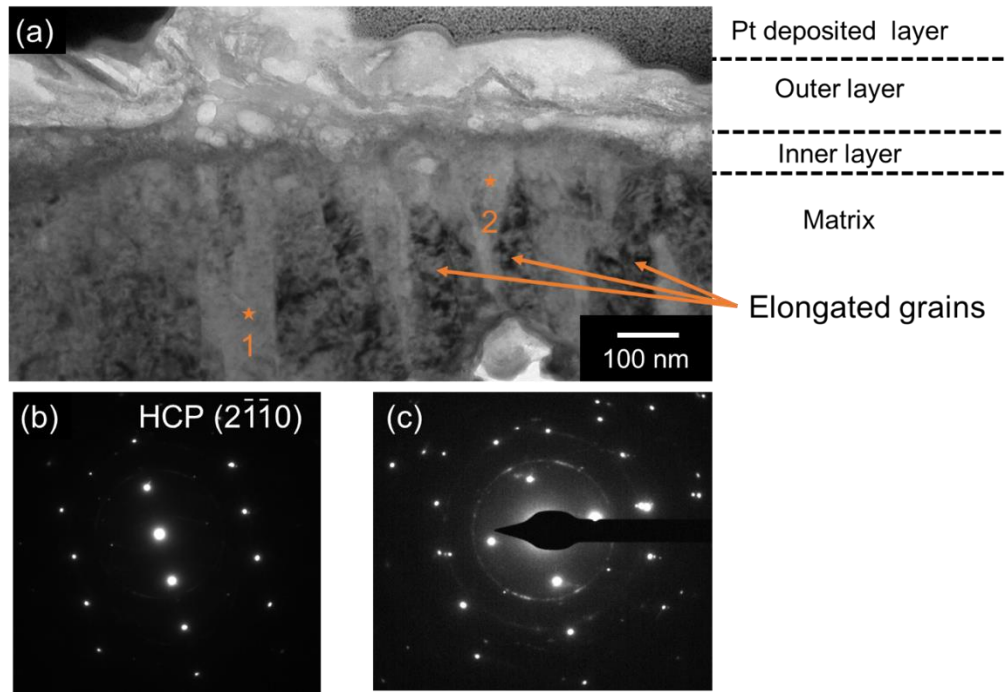

Fig. 2. (a) A bright field TEM image of the cross-sectional surface lamella from the Mg-Li(-Al-Y-Zr) alloy following 24 h of immersion in 0.1 M NaCl and exposure to air for 7 days. (b) and (c) reveal the SAED patterns collected from the positions marked as "1" and "2" in (a), respectively.

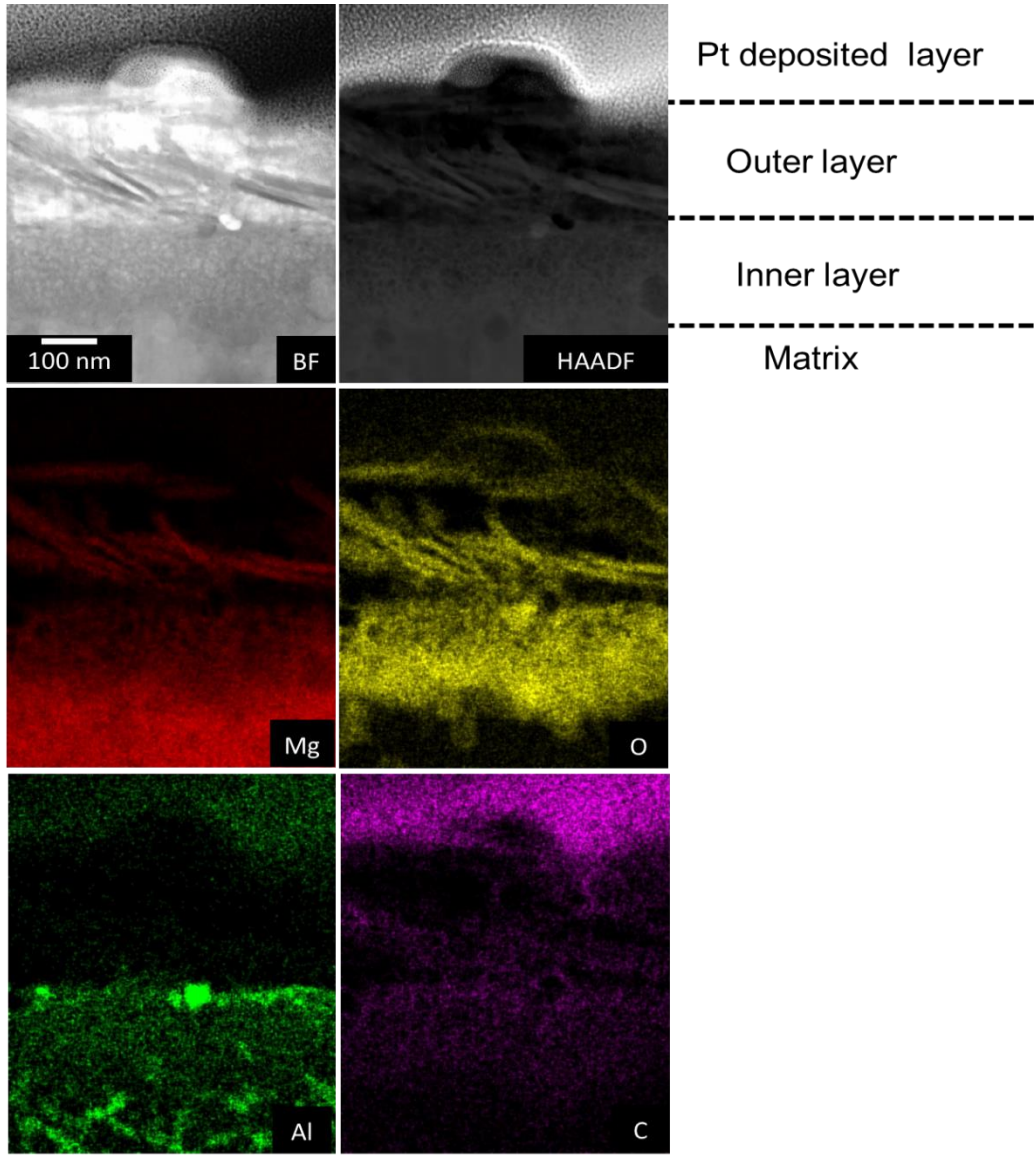

Fig. 3. Bright field and high-angle annular dark-field STEM images, along with the corresponding EDXS elemental maps of Mg, Al, C and O from the cross-sectional surface lamella from the Mg-Li(-Al-Y-Zr) alloy following 24 h of immersion in 0.1 M NaCl and exposure to air for 7 days.

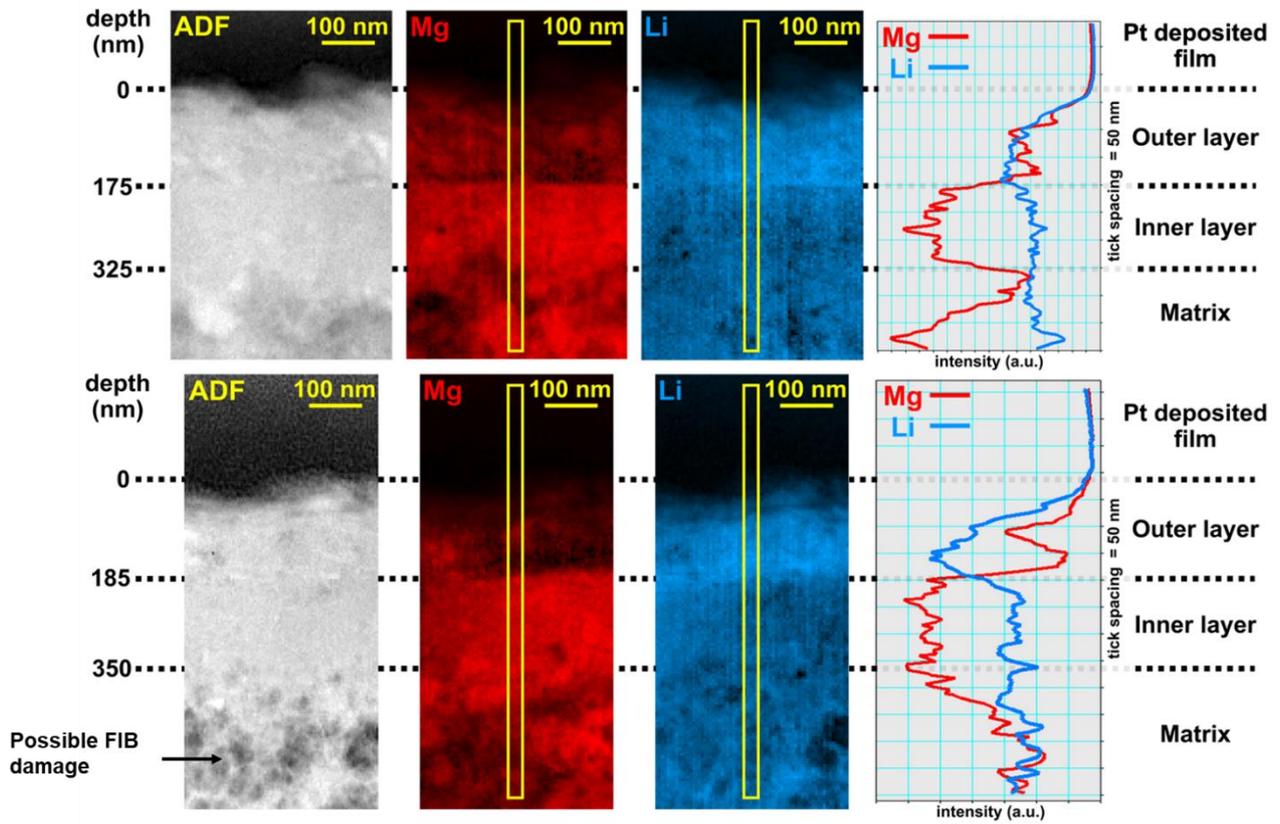

Fig. 4. Two separate regions shown in the annular dark-field images (ADF) from different regions of the TEM lamella specimen (for the same conditions in Fig. 3) were mapped for Mg (red) and Li (blue) content using the spectrum imaging method described in §2. The vertically elongated yellow boxes show the loci along which the intensity profiles shown at right were taken. The horizontal width of each box indicates the spatial integration width used to smooth each intensity profile. There are four distinct regions that make up the scanned volumes of material and these are marked as the "Pt deposited film", the "outer layer", the "inner layer" and the "matrix" where the latter three are part of the alloy surface region. Approximate depths from the top surface of the alloy are indicated at left.